\documentclass{article} 
\usepackage{amsmath,amssymb,amsbsy,amsfonts,comment}
\makeatletter

          \@addtoreset{equation}{section}
       \makeatother
\newcommand{\be}{\begin{equation}}
\newcommand{\ee}{\end{equation}}
\allowdisplaybreaks 

\newenvironment{namelist}[1]{%
\begin{list}{}
{
\settowidth{\labelwidth}{#1}
\setlength{\leftmargin}{1.1\labelwidth}}
}{%
\end{list}}
\begin{document}
\title{\bf Noncommutative gauge theory with the symmetry
$\boldsymbol{U(n\otimes m)}\ast $ and
standard-like model with fractional charges}%

\author{Yoshitaka {\sc Okumura}\\
{Department of Natural Science, 
Chubu University, Kasugai, {487-8501}, Japan}
}%
\date{}%
\maketitle
\begin{abstract}
$\text{U}(n\otimes m)\ast$ gauge field theory  on noncommutative spacetime
is formulated and the standard-like model with the symmetry ${\text{U}(3_c\otimes 2\otimes 1_{\text{\scriptsize$Y$}})\ast}$ is reconstructed based on it.
On the noncommutative spacetime, 
the representation that  fields belong to is fundamental, adjoint or bi-fundamental. For this reason, one had to construct the standard model by use of 
bi-fundamental representations. However, we can reconstruct the standard-like model with only fundamental and adjoint representation and without using  bi-fundamental representations.
It is well known that the charge of fermion is 0 or $\pm1$ in the
U(1) gauge theory on noncommutative spacetime. 
Thus, there may be 
no room to incorporate the noncommutative U(1) gauge theory 
into the standard model because the quarks have fractional charges.
However, 
it is shown in this article   that
there is the noncommutative gauge theory with arbitrary charges
which symmetry is for example $\text{U}(3\otimes 1)\ast$.
This type of gauge theory emerges from 
the spontaneous breakdown of 
 the noncommutative  U(4)$\ast$ gauge theory 
in which 
the gauge field contains the 0 component $A_\mu^0(x,\theta)$.
The standard-like model in this paper also has fermion fields with fractional charges.
Thus, 
the noncommutative gauge theory with fractional U(1) charges 
can not exist alone, but it must coexist with 
noncommutative nonabelian gauge theory.
\end{abstract}



\thispagestyle{empty}
\section{Introduction\label{S1}}
In the past several years, field theories on the noncommutative(NC) spacetime
have been extensively studied from many different aspects. The motivation 
comes from the string theory which makes obvious that end points of 
the open strings trapped on the D-brane in the presence of 
 two form B-field background turn out to be noncommutative 
\cite{SJ}
and 
then the noncommutative supersymmetric gauge theories appear as the low energy effective theory of such D-brane \cite{DH}, \cite{SW}.
\par
After the completion of standard model in particle physics more than 
decades ago, there have been several occasions  that indicated
the experimental deviations from the standard model. However, 
such deviations ultimately shrank to nothing and the correctness of 
the standard model has been confirmed. Thus,  many advanced 
theories beyond standard model must reduce to the standard model in their characteristic limits. If not, such theory is branded not to be the qualified theory beyond standard model. This is the case also in the NC field theory.
However, there are many difficulties in  gauge theories on NC spacetime
because of the noncommutativity of spacetime. One of the difficulty 
\cite{Gracia, Terashima,Chaichian}
is that
in order to respect the gauge invariance the field must belongs to the fundamental, adjoint or bi-fundamental representation of the gauge group 
on NC spacetime. 
In the standard model, three symmetries such as color, weak isospin and 
hypercharge are there. Thus, quarks have to be bi-fundamental
in the symmetry $\text{U}(3)_c\times\text{U}(2)_L\times\text{U}(1)_Y$
as in \cite{CPST,Khoze}
which tried to construct the standard model on NC spacetime. 
\par
In this article, 
$\text{U}(n\otimes m)\ast$ gauge field theory  on noncommutative spacetime
is formulated.
$\text{U}(n\otimes m)\ast$ gauge group  is the resultant gauge group from 
the spontaneous symmetry breakdown of ${\text{U}}(N)\ast\,(N=n+m)$.
It reduces to $\text{U}(n\otimes m)$  
on the commutative spacetime which is not ${\text{U}}(N)$ but isomorphic to 
$\text{SU}(n)\otimes\text{SU}(m)\otimes \text{U}(1)$.
The standard-like model with
the symmetry ${\text{U}(3_c\otimes 2\otimes 1_{\text{\scriptsize$Y$}})\ast}$  is reconstructed 
without using the bi-fundamental representation of fields.
It should be noted that 
the symmetry ${\text{U}(3_c\otimes 2\otimes 1_{\text{\scriptsize$Y$}})\ast}$ is different from $\text{U}(3)_c\ast\times\text{U}(2)_L\ast\times\text{U}(1)_Y\ast$ as explained in the fourth section.
\par
There are other difficulties in NC gauge theory.
Hayakawa \cite{Hayakawa} indicated that the matter field must have charge 
0 or $\pm1$ in U(1) NC  gauge theory  in order to keep the gauge invariance of the theory.
Matsubara \cite{Matsubara} and Armoni \cite{Armoni} also indicated that U(N) gauge theory has the consistency in calculations of gluon propagator and  three gluons vertex to one loop order, whereas SU(N) gauge theory is not consistent. 
These problems make it very difficult to reproduce the standard model in the framework of NC gauge theory.
The indication made by Hayakawa \cite{Hayakawa} is serious since 
quarks in standard model have fractional charges. There must be 
the story other than Hayakawa's indication if the noncommutativity
on the spacetime is somewhat true in nature. 
We can overcome this difficulty by
considering nonabelian ${\text{U}}(n\otimes m)\ast$ gauge theory \cite{Okumura} which results from
the spontaneous symmetry breakdown of ${\text{U}}(N)\ast$ gauge theory.
We conclude that  gauge theory that has the matter fields with fractional charges can not exist alone, but it must coexist with 
NC nonabelian gauge theory.\par
This article consists of 5 sections. In second section, 
the NC nonabelian gauge theory with the symmetry U(n $\otimes$  m)$\ast$
 is proposed.  
In third section, the spontaneous symmetry breakdown of SU(4)$\ast$ gauge
theory is discussed in order to show that quarks has the fractional 
$B$-charges whereas the lepton has charge $-1$. 
In fourth section, the standard-like model with the symmetry ${\text{U}(3_c\otimes 2\otimes 1_{\text{\scriptsize$Y$}})\ast}$ is proposed
 to show that fields with color as well as flavor
quantum numbers can be expressed in terms of the fundamental 
representation and also fields with fractional charges are  incorporated
in the gauge field on NC spacetime.
 The last section is devoted to discussions and conclusions.

\section{Nonabelian gauge theory on noncommutative spacetime\label{S2}}
Let us first consider the nonabelian gauge theory 
on the NC spacetime
with the symmetry $\text{U}(n+m)\ast$  given by the Lagrangian
\begin{align}
{\cal L}=&-\frac12\text{Tr}\left[F_{\mu\nu}(x)\ast F^{\mu\nu}(x)\right]
+{\mit\bar\Psi}(x)\ast\{i\gamma^\mu(\partial_\mu-igA_\mu(x))-m\}
\ast{\mit\Psi}(x)\nonumber\\
&+\text{Tr}\left[({\cal D}^\mu\varphi(x))^\dag\ast {\cal D}_\mu\varphi(x)
+m^2\varphi(x)^\dag \ast\varphi(x)-\lambda (\varphi(x)^\dag \ast\varphi(x))^2
\right],
\label{2.1}
\end{align}
where we omit the gauge fixing and FP ghost terms. 
The Moyal star product of functions $f(x)$ and $g(x)$ is 
defined as
\begin{equation}
f(x)\ast g(x)=\left.e^{\frac{i}{2}\theta^{\mu\nu}\partial^1_\mu\partial^2_\nu}f(x_1)g(x_2)\right|_{x_1=x_2=x},
\end{equation}
where $\theta^{\mu\nu}$ is constant with two subscripts 
to characterize the noncommutativity of
spacetime.  The noncommutative parameter  $\theta^{\mu\nu}$ is usually seemed to be a constant not to transform corresponding to Lorentz transformation.
$\mit\Psi(x)$ is the fermion field with the fundamental representation.
The quantity
\begin{equation}
F_{\mu\nu}(x)=\partial_\mu A_\nu(x)-\partial_\nu A_\mu(x)-ig\,[A_\mu(x),\,A_\nu(x)]_*\label{2.2}
\end{equation}
is the field strength of gauge field 
with the configuration
\begin{equation}
A_\mu(x)=\sum_{a=0}^{n^2-1}\sum_{b=0}^{m^2-1}A^{ab}_\mu(x)T^a\otimes {\hat T}^b.\label{2.3}
\end{equation}
and the field  $\varphi(x)$ is the Higgs boson belonging to the adjoint or fundamental representation of $\text{U}(n)\otimes\text{U}(m)$ which covariant derivative is  
\begin{equation}
{\cal D}^\mu\varphi(x)=\partial^\mu\varphi(x)-ig[A^\mu(x),\,\varphi(x)]_\ast,
\end{equation}
or
\begin{equation}
{\cal D}^\mu\varphi(x)=\partial^\mu\varphi(x)-igA^\mu(x)\ast \varphi(x).
\end{equation}
\par
\par
 The gauge transformations of fields in \eqref{2.1} are defined as
\begin{align}
\begin{aligned}
& A^g_\mu(x)=U(x,\theta)\ast A_\mu(x)\ast U^{-1}(x,\theta)+\frac{i}{g}U(x,\theta)\ast\partial_\mu U^{-1}(x,\theta),\\
&{\mit\Psi}^g(x)=U(x,\theta)\ast{\mit\Psi}(x),\\
&\varphi^{\,g}(x)=U(x,\theta)\ast \varphi(x)\ast U^{-1}(x,\theta)
\label{2.6}
\end{aligned}
\end{align}
where the gauge transformation function $U(x,\theta)$ is written as
\begin{equation}
U(x,\theta)=e^{i\alpha(x,\theta)\ast}=\sum_{n=0}^\infty \frac{i^n}{n!}\;\alpha(x,\theta)\ast\alpha(x,\theta)\ast\alpha(x,\theta)\ast\cdots\ast\alpha(x,\theta)
\label{2.9}
\end{equation}
in terms of the Lie algebra valued function
\begin{equation}
\alpha(x,\theta)=\sum_{a=0}^{n^2-1}\sum_{b=0}^{m^2-1}\alpha^{ab}(x,\theta)\,T^a\otimes {\hat T}^b\label{2.10}
\end{equation}
with the condition that
\begin{equation}
\alpha^{ab}(x,\theta)=f^a(x,\theta)\ast g^b(x,\theta).
\end{equation}
The ensemble of gauge function \eqref{2.9} is
extended gauge group denoted by $\text{U}(n\otimes m)\ast$.
The star commutator between two Lie algebra valued functions 
is calculated as
\begin{align}
[\alpha(x,\theta),\,\beta(x,\theta)]_{\ast}=&\sum_{a,c=0}^{n^2-1}\sum_{b,d=0}^{m^2-1}\left(\alpha^{ab}(x,\theta)\ast\beta^{cd}(x,\theta)T^aT^c\otimes {\hat T}^b{\hat T}^d-\beta^{cd}(x,\theta)\ast\alpha^{ab}(x,\theta)T^cT^a\otimes {\hat T}^d {\hat T}^b\right).
\label{2.12a}
\end{align}
Thus,  the enveloping
Lie algebra closes within itself for the star commutator \eqref{2.12a}.
It should be noted that 
\begin{equation}
\lim_{\theta\to0}U(x,\theta)\ast=\exp\left[{i\sum_{a=0}^{(n^2-1)} f^a(x)T^a}\right]
\otimes \exp\left[{i\sum_{b=0}^{(m^2-1)} g^b(x){\hat T}^b}\right]
\in \text{SU}(n)\otimes \text{SU}(m)\otimes\text{SU}(1),\label{2.14}
\end{equation}
which indicates that the extended group $\text{U}(n\otimes m)\ast$ reduces to nonabelian group $\text{SU}(n)\otimes\text{SU}(m)\otimes\text{SU}(1)$ when $\theta^{\mu\nu}$ approaches to 0. However,
 we consider this limit in classical level, not in quantum level.
\par
Under the gauge transformation in \eqref{2.6} 
the field strength $F_{\mu\nu}(x)$ and
the covariant derivative of ${\cal D}_\mu\varphi(x)$
are transformed covariantly
\begin{align}
&F_{\mu\nu}^g(x)=U(x,\theta)\ast F_{\mu\nu}(x)\ast U^{-1}(x,\theta),\\
&({\cal D}_\mu\varphi(x))^g=U(x,\theta)\ast 
{\cal D}_\mu\varphi(x)\ast U^{-1}(x,\theta).\label{2.16}
\end{align}
Then, the gauge field term in \eqref{2.1} is transformed as in
\begin{equation}
\text{Tr}\left[F_{\mu\nu}^g(x)\ast {F^g}^{\mu\nu}(x)\right]
=\text{Tr}\left[U(x,\theta)\ast F_{\mu\nu}(x)\ast F^{\mu\nu}(x)
\ast U^{-1}(x,\theta)\right]
\label{2.22a}
\end{equation}
which shows the gauge term itself is not gauge invariant because of the 
Moyal $\ast$product but the action is invariant thanks to the rule
\begin{equation}
\int d^4x \;f(x)\ast g(x)=\int d^4x \;g(x)\ast f(x).
\end{equation}
We call this situation 
 pre-invariance to gauge transformation. That is, 
the gauge boson term in Lagrangian is pre-invariant. 
That is the case for the Higgs boson term in \eqref{2.1} because 
 of \eqref{2.16}. 
On the other hand, the fermion term in Eq.\eqref{2.1} is invariant under gauge transformations \eqref{2.6}. 
\par
Here, we take $N=4$ without loss of generality in order to define the
more general gauge theory on NC spacetime.
As stated in the next section, we obtain the $\text{U}(3\otimes 1)\ast$ gauge theory resulting from the spontaneous symmetry breakdown of 
$\text{U}(4)\ast$ gauge theory. However, regardless of the spontaneous breakdown, we can consider the $\text{U}(3\otimes 1)\ast$ gauge theory  as the  gauge theory with the generators
${\lambda'}^a$ which is $4\times 4$ matrix 
constructed from the Gell-Mann matrix $\lambda^a$ by adding $0$ components 
to fourth line and column
\be
   {\lambda'}^a  =\begin{pmatrix}  & &  & 0\cr
                                & \lambda^a &  & 0\cr
                                &  &   &0\cr
                              0 & 0 & 0 & 0 \cr
\end{pmatrix}
     \label{2.17}
\ee
, $\lambda^{15}=\frac{1}{2\sqrt{6}}\text{Diag}(1,1,1,-3)$ and $4\times 4$ unit matrix
$\lambda^0$.
\par
It should be noted that the $\text{U}(3\otimes 1)\ast$ 
gauge theory is different with
the product gauge theory with the symmetry 
$\text{U}(3)\times\text{U}(1)\ast$ because of the spacetime noncommutativity.
We can also define the more general gauge theory with the symmetry such as $\text{U}(3\otimes 2\otimes 1)\ast$ consisting of the $16\times16$ matrix generators, which is used to reconstruct the standard-like model in the fourth section. It should be noticed that the noncommutative gauge group 
$\text{U}(3\otimes 2\otimes 1)\ast$ becomes 
$\text{U}(3)\times\text{U}(2)\times\text{U}(1))\ast$ except for the overall
$\text{U}(1))$ in the commutative limit though $\text{U}(N)\ast$ would not resolve into $\text{SU}(N)\times\text{U}(1)$ in the quantum level.
\section{The spontaneous breakdown of U(4)$\ast$ gauge theory}
SO(10) grand unified theory (GUT) including its supersymmetric version 
is most promising model in particle physics 
since it can incorporate the 15 existing fermions in addition to 
the right-handed neutrino and has possibilities to explain so many
phenomenological puzzles. Pati-Salam symmetry $\text{SU(4)}\times \text{SU(2)}_L\times \text{SU(2)}_R$ is one of the intermediate symmetry of the spontaneous breakdown of SO(10) GUT. This symmetry spontaneously breaks down to
the left-right symmetric gauge model with the symmetry 
$\text{SU(3)}_c\times \text{SU(2)}_L\times \text{SU(2)}_R
\times \text{U(1)}_B$.
In this stage, the spontaneous breakdown 
$\text{SU(4)}\to \text{SU(3)}_c\times \text{U(1)}_B$
occurs. $B$ charge of fermions is given by
\begin{equation}
Q_B
\begin{pmatrix}
q^r \\ q^g \\ q^b \\ l
\end{pmatrix}
=
\begin{pmatrix}
\frac13 & 0 & 0 & 0\\ 0 & \frac13 & 0 & 0 \\ 
0 & 0 & \frac13 & 0 \\  0 & 0 & 0 & \text{\small$-{1}$}
\end{pmatrix}
\begin{pmatrix}
q^r \\ q^g \\ q^b \\ l
\end{pmatrix}.\label{3.1}
\end{equation}
As an example, we pick up this process in order to investigate
whether fermions have $B$ charge in \eqref{3.1} in the noncommutative 
version of SU(4) gauge theory.
\par
The gauge boson in SU(4)$\ast$ gauge theory is expressed 
in terms of 16 component gauge bosons by
\begin{align}
A_\mu(x)&=\sum_{a=0}^{15}A_\mu^a(x)\,T^{\,a}\nonumber\\[2mm]
&=\frac{1}{{2}}\begin{pmatrix}
A_\mu^{11}& G^{1}_\mu & G^{2}_\mu & X^{1}_\mu \\[2mm]
{\bar G}^{1}_\mu & A_\mu^{22}
 & G^{3}_\mu & X^{2}_\mu \\[2mm]
{\bar G}^{2}_\mu & {\bar G}^{3}_\mu & A_\mu^{33} & X^{3}_\mu \\[2mm]
{\bar X}^{1}_\mu & {\bar X}^{2}_\mu & {\bar X}^{3}_\mu & 
A_\mu^{44}\\
\end{pmatrix}.
\end{align}
where
\begin{align}
&\left\{\begin{aligned}
&A_\mu^{11}=\dfrac{1}{\sqrt{2}}A^0_\mu+A^{3}_\mu
+\dfrac{1}{\sqrt{3}}A^{8}_\mu+\dfrac{1}{\sqrt{6}}A^{15}_\mu ,\\
&A_\mu^{22}=\dfrac{1}{\sqrt{2}}A^0_\mu
-A^{3}_\mu+\dfrac{1}{\sqrt{3}}A^{8}_\mu+\dfrac{1}{\sqrt{6}}A^{15}_\mu\\
&A_\mu^{33}=\dfrac{1}{\sqrt{2}}A^0_\mu
-\dfrac{2}{\sqrt{3}}A^{8}_\mu
+\dfrac{1}{\sqrt{6}}A^{15}_\mu\\
&A_\mu^{44}=\dfrac{1}{\sqrt{2}}A^0_\mu-\dfrac{3}{\sqrt{6}}A^{15}_\mu
\end{aligned}\right.\\
&G_\mu^1=A_\mu^1-iA_\mu^2,\hskip5mm G_\mu^2=A_\mu^4-iA_\mu^5,\hskip5mm
G_\mu^3=A_\mu^6-iA_\mu^7,\\
&X_\mu^1=A_\mu^9-iA_\mu^{10},\hskip5mm X_\mu^2=A_\mu^{11}-iA_\mu^{12},\hskip5mm
X_\mu^3=A_\mu^{13}-iA_\mu^{14}.
\end{align}
The gauge field $A_\mu(x)$ 
contains 8 color gluons, 6 gauge bosons causing proton decay,  
one extra boson $A^{15}_\mu(x)$, and one 0 component boson $A^{0}_\mu(x)$ dependent on other bosons. Here, we denote $A^{15}_\mu(x)$ by $B_\mu(x)$ and call it 
$B$-field.

The vacuum expectation value of Higgs boson $\varphi(x)$ takes the form
\begin{equation}
<\varphi(x)>=v
\begin{pmatrix}
1 & 0 & 0 & 0 \\
0 & 1 & 0 & 0 \\
0 & 0 & 1 & 0 \\
0 & 0 & 0 & -3 
\end{pmatrix}
\label{3.3}
\end{equation}
which yields the gauge boson mass term
\begin{align}
\left|-ig[A_\mu(x),\,<\varphi(x)>]\rule{0mm}{3.3mm}\right|^2&=
{8g^2v^2}\left({X^1}\rule{0mm}{2mm}^\mu {\bar X}_\mu^1
+{X^2}\rule{0mm}{2mm}^\mu {\bar X}_\mu^2
+{X^3}\rule{0mm}{2mm}^\mu {\bar X}_\mu^3\right).\label{3.4}
\end{align}
Equation \eqref{3.4} shows that 6 proton decay causing gauge bosons acquire
mass, so that symmetries relating to Lie algebra $T^a\;(a=1,2,\cdots,8,15)$
keep unbroken. Let us consider the gauge transformation specified by
\begin{align}
U^{cb}(x,\theta)=e^{i\alpha(x,\theta)\ast}
=\exp i\left\{{\sum_{a=0}^8T^a\alpha^a(x,\theta)\ast+
T^{15}\alpha^{15}(x,\theta)\ast}\right\}.
\label{3.5}
\end{align}
Under this gauge transformation,
the part of the gauge field $A_\mu(x)$ in \eqref{2.3}
\begin{align}
A_\mu^{cb}(x)={\sum_{a=0}^8A^a_\mu(x)\,T^a+
B_\mu(x)T^{15}}
\end{align}
and fermion field $\mit\Psi(x)$ and Higgs field transform in the similar way as in \eqref{2.6}.
Thus, it is easily shown that
the Lagrangian \eqref{2.1} is still pre-invariant after the spontaneous
breakdown resulting from \eqref{3.3}.
This indicates that color symmetry yielding the strong interaction and 
$B$-symmetry due to the generator $T^{15}$ remain unbroken.
In the commutative field theory, this breakdown is written as
\begin{align}
\text{SU(4)} \rightarrow \text{SU(3)}\times \text{U(1)}.
\end{align}
However, we can't do it in the same way because 
\begin{align}
U^{cb}(x,\theta)\ne 
\exp i\left\{\sum_{a=1}^8\alpha^a(x,\theta)\ast\,T^a\right\}
\ast
\exp i\left\{{
\alpha^{15}(x,\theta)\ast\,T^{15}}\right\}
\label{3.8}
\end{align}
owing to the Moyal product. Thus, in the NC field theory,
we should write the spontaneous breakdown explained so far as
\begin{align}
\text{U(4)}\ast \rightarrow \text{U(3$\,\otimes $1)}\ast.
\end{align}

\par
Interaction terms between fermion and $B$-gauge field extracted from the fermion term in \eqref{2.1} is given by
\begin{align}
{\cal I}_D&={\mit\bar\Psi}(x)\ast\{\gamma^\mu(gB_\mu(x)T^{15})\}
\ast{\mit\Psi}(x)\nonumber\\
&=\frac{3}{2\sqrt{6}}\,g\,{\mit\bar\Psi}(x)\ast\gamma^\mu
\begin{pmatrix}
\frac13 & 0 & 0 & 0 \\
0 & \frac13 & 0 & 0 \\
0 & 0 & \frac13 & 0 \\
0 & 0 & 0 & \text{\small$-1$} 
\end{pmatrix}
B_\mu(x)
\ast{\mit\Psi}(x)
\end{align}
Then, if we define $B$-charge operator $Q_B$ 
and $B$-charge $e_{\text{\tiny$B$}}$
\begin{align}
Q_B=\frac{2\sqrt{6}}{3}T^{15},\hskip1cm 
e_{\text{\tiny$B$}}=\frac{3}{2\sqrt{6}}\,g
\end{align}
\eqref{3.1} is reproduced.\par
We considered the spontaneous breakdown
of U(4)$\ast$ gauge symmetry down to U(3$\,\otimes\,$1)$\ast$ symmetry. 
Thus, charges of fermions are limited as shown in \eqref{3.1}.
However, apart from such construction, we can considered  such a case that 
if the Lagrangian is pre-invariant under the gauge transformation function $U^s(x,\theta)$ given by
\begin{align}
U^{s}(x,\theta)=e^{i\alpha(x,\theta)\ast}
=\exp i\left\{\sum_{a=0}^8\,T^a\,\alpha^a(x,\theta)\ast+
Q\,\beta(x,\theta)\ast\right\}
\label{3.1f}
\end{align}
where
\begin{align}
Q=
\begin{pmatrix}
e & 0 & 0 & 0 \\
0 & e & 0 & 0 \\
0 & 0 & e & 0 \\
0 & 0 & 0 & e' 
\end{pmatrix}
\end{align}
with arbitrary constants $e$ and $e'$, fermions may have arbitrary 
charges. This is because 
Interaction terms between fermion and $B$-gauge is given by
\begin{align}
{\cal I}_D&={\mit\bar\Psi}(x)\ast\{\gamma^\mu(gB_\mu(x)Q)\}
\ast{\mit\Psi}(x).
\end{align}
\par
If there is only U(1)$\ast$ gauge symmetry, the gauge transformation 
of gauge field 
$A_\mu(x)=Q B_\mu(x)$
given by
\begin{align}
U^{e}(x,\theta)
=\exp i\left\{
Q\,\beta(x,\theta)\ast\right\}.
\label{4.3}
\end{align}
leads to the inconsistency as indicated by Hayakawa \cite{Hayakawa}. 
But, in our case, there are two kinds of symmetry and therefore, 
the gauge transformation of gauge field 
$A_\mu(x)=\sum_{a=0}^8T^a{A^a}_\mu(x,\theta)+
Q B_\mu(x)$ given by \eqref{3.1f} has nothing to do with any contradiction because of $A_\mu^0(x,\theta)$ existence.
\section{The standard-like model with the symmetry $\boldsymbol{\text{U}(3_c\otimes 2\otimes 1_{\text{\scriptsize$Y$}})\ast}$}
We construct the standard-like model to show that fields with color as well as flavor
quantum numbers can be expressed in terms of the fundamental 
representation and also fields with fractional charges are  incorporated
in the gauge field on NC spacetime.\par
We explain the gauge group
${\text{U}(3_c\otimes 2\otimes 1_{\text{\scriptsize$Y$}})\ast}$
which generators are formed by the $8\times8$ matrices.
The generators of color sector are 
\be
\Gamma^a=1^2\otimes {\lambda'}^a\,(a=1,\cdots,8)\label{4.1}
\ee where $1^4$
is the 4 dimensional unit matrix and ${\lambda'}^a$ is given in \eqref{2.17}.
The generators of the weak isospin sector are written as
\be
\Gamma^i=
\tau^i\otimes1^4 
\ee
where $\tau^i\,(i=1,2,3)$ is the Pauli matrix.
The generator of the hypercharge is the 8 dimensional diagonal matrix
\be
\Gamma^{\text{\scriptsize{$Y$}}}={\rm Diag}\left(\frac13,\frac13,\frac13,-1,\frac13,\frac13,\frac13,-1\right)
\ee
It should be noted that all these generators  are  8 dimensional matrices
and form closed algebra.
The whole abelian parts are denoted by $\Gamma^0$
which are not explicitly written.
With these generators explicitly written above, the group element
of 
${\text{U}(3_c\otimes 2\otimes 1_{\text{\scriptsize$Y$}})\ast}$
is denoted as
\begin{align}
g(x)=\text{Exp}\left\{\,-\frac{i}{2}\,\big(\alpha^0(x)\Gamma^0+\alpha^a(x)\Gamma^a+\alpha^i(x)\Gamma^i+\alpha^{\text{\scriptsize{$Y$}}}(x)\Gamma^{\text{\scriptsize{$Y$}}}\right\}_\ast.\label{4.6}
\end{align}
It should be noted that $g(x)$ is not factored out into $2\otimes4\otimes1$ matrix owing to the spacetime noncommutativity.
\par
The aggregate gauge field $A_\mu$ in terms of all gauge fields
is described in the equation
\begin{align}
{\cal A}_\mu(x)=\frac12\left(
g_0A_\mu^0(x)\Gamma^0\right.&\left.+g_cG_\mu^a(x)\Gamma^a+g{A}_\mu^i(x)\Gamma^i+g'B_\mu(x)\Gamma^{\text{\scriptsize{$Y$}}}\right),\label{4.5qw}
\end{align}
where $g_0,\,g_c,\,g,\,g'$ are the coupling constants corresponding with each
gauge field. Field strength $\cal{F_{\mu\nu}}$ is defined as
\begin{align}
\begin{aligned}
{\cal{F_{\mu\nu}}}(x)&=\partial_\mu{\cal A}_\nu(x)-\partial_\nu{\cal A}_\mu(x)
+[{\cal A}_\mu(x),\;{\cal A}_\nu(x)]_\ast\\
&=\frac12\left(g_0F_{\mu\nu}^0(x)\Gamma^0+g_cG_{\mu\nu}^a(x)\Gamma^a+g{F}_{\mu\nu}^i(x)\Gamma^i+g'B_{\mu\nu}(x)\Gamma^{\text{\scriptsize{$Y$}}}\right)\\&\text{\quad}
+\text{terms resulting from the noncommutativity of spacetime},
\end{aligned}
\end{align}
where
\begin{align}
F_{\mu\nu}^0(x)&=\partial_\mu{A}_\nu^0(x)-\partial_\nu{A}_\mu^0(x),\\
G_{\mu\nu}^a(x)&=\partial_\mu{G}_\nu^a(x)-\partial_\nu{G}_\mu^a(x)
+g_cf^{abc}{G}_\mu^b(x)\ast{G}_\nu^c(x),\\
{F}_{\mu\nu}^i(x)&=\partial_\mu{A}_\nu^i(x)-\partial_\nu{A}_\mu^i(x)
+g\epsilon^{ijk}{A}_\mu^j(x)\ast{A}_\nu^k(x),\\
B_{\mu\nu}(x)&=\partial_\mu B_\nu(x)-\partial_\nu B_\mu(x).
\end{align}
The gauge field $A_\mu(x)$ transforms under the gauge transformation \eqref{4.6} by
\be
{\cal A}_\mu^{'}(x)=g(x)\ast {\cal A}_\mu(x)\ast g^{-1}(x)+ig(x)\ast\partial_\mu g^{-1}(x).\label{4.13}
\ee
Then, the field strength ${\cal{F_{\mu\nu}}}(x)$
is transformed covariantly
\be
{\cal{F_{\mu\nu}}}'(x)=g(x)\ast{\cal{F_{\mu\nu}}}(x)\ast g^{-1}(x).
\ee

We find the Yang-Mills Lagrangian $L_{\text{\tiny{$YM$}}}$ 
\be
L_{\text{\tiny{$YM$}}}=-\frac{1}{k^2}\text{Tr}\,\left[\,
{\cal{F_{\mu\nu}}}(x)\ast{\cal{F^{\mu\nu}}}(x)\,\right]
\ee
from which the relationships of coupling constants
\begin{align}
 g_c^2=\frac{1}{4}k^2, && g^2=\frac{1}{4}k^2,&&
{g'}^2=\frac{3}{8}k^2\label{4.14}
\end{align}
are obtained.
It should be noted that the integral of $L_{\text{\tiny{$YM$}}}$ over $x$ is gauge invariant.
\par
Let us represent the  fermion field $\psi$ as a 8 dimensional vector 
with borrowing the names of existing leptons and quarks 
and then  give the correct Dirac Lagrangian for the fermion sector 
in the standard-like model. Hereafter, the argument $x$ is often abbreviated 
if there is no confusion. 
\be
\psi=\begin{pmatrix} 
                                u^r\\[1mm]
                                u^g\\[1mm]
                              u^b\\[1mm]
                                \nu\\[1mm]
                                d^r\\[1mm]
                                d^g\\[1mm]
                                d^b\\[1mm]
                                e\\[1mm] 
                            \end{pmatrix}
\label{4.17}
\ee
with color indices $r$, $g$ and $b$. 
According to \eqref{4.6}, the gauge transformation of fermion field $\psi$ in \eqref{4.17} is given by
\be
\psi'=g(x)\ast \psi. \label{4.19}
\ee
\par
With this configuration of fermion field, the Dirac Lagrangian is simply written as
\be
L_{\text{\tiny{$D$}}}=i{\bar\psi}\ast \gamma^\mu\left(\partial_\mu-i{\cal A}_\mu\right)\ast \psi.\label{4.20}
\ee
It is evident that the Dirac Lagrangian $L_{\text{\tiny{$D$}}}$ is gauge invariant according to
\eqref{4.13} and \eqref{4.19}.
\par
We consider a Higgs field ${\mit\Phi}$ 
 which belong to the adjoint representation in similar way as in \eqref{4.5qw}.
 Gauge transformation of ${\mit\Phi}$ is subject to 
\be
{\mit\Phi}'=g(x)\ast{\mit\Phi}\ast g^{-1}(x).\label{4.23}
\ee
Equation \eqref{4.23} yields the covariant derivative of Higgs field 
\be
{\cal D}_\mu{\mit\Phi}_i=\partial_\mu{\mit\Phi}_i-i[{\cal A}_\mu,\,{\mit\Phi}
]_\ast
\ee
from which we can construct the Higgs-gauge interaction term
\begin{align}
{\cal L}_{\text{\tiny{$D$}}}=\frac1{{k'}^2}{\rm Tr}\big({\cal D}_\mu{\mit\Phi}\big)^\dag \ast \big({\cal D}^\mu{\mit\Phi}\big)
\end{align}
with the normalization factor $k'$.
\par
Yukawa interaction between fermion and Higgs fields is given as
\be
{\cal L}_{\text{\tiny{$Y$}}}=g_{\text{\tiny{$Y$}}}
{\bar\psi}\ast {\mit\Phi}\ast \psi
\label{4.26}
\ee
where $g_{\text{\tiny{$Y$}}}$ is the Yukawa coupling matrix.
It should be noted that  
${\cal L}_{\text{\tiny{$Y$}}}$ in \eqref{4.26} is gauge invariant.
\par
Spontaneous breakdown of gauge symmetry is caused by
the vacuum expectation value of Higgs boson
\be
<{\mit\Phi}>=\begin{pmatrix} 1 & 0 \\
                             0 & -1 \end{pmatrix}\otimes1^4\mu
\ee
It is obvious that $[\,\Gamma^a,\,<{\mit\Phi}>]=0$, and therefore, the color symmetry does not  break spontaneously. Then, the gauge symmetry 
${\text{U}(3_c\otimes 2\otimes 1_{\text{\scriptsize$Y$}})\ast}$ spontaneously breaks down to 
${\text{U}(3_c\otimes 1_{\text{em}})\ast}$.
Thus, gluon mass remains zero.
In addition, since $[\,\Gamma^0,\,<{\mit\Phi}>]=0$, $[\,\Gamma^3,\,<{\mit\Phi}>]=0$ and
$[\,\Gamma^{\text{\tiny{$Y$}}},\,<{\mit\Phi}>]=0$, 
gauge bosons with respect to these relations remain massless.
However, charged gauge bosons acquire the masses due to the symmetry breakdown.
\begin{align}
\text{Gauge boson mass}^2=&\text{Tr}[\,i{\cal A}_\mu,\,<{\mit\Phi}>]^2=
-2{g^2}\text{Tr}\begin{pmatrix}
0^4 &-{W}_\mu^-  \\
{W}_\mu^+ & 0^4
\end{pmatrix}^2\mu^2=
m_{\text{\tiny{$W$}}}^2W_\mu^+{W^\mu}^{-},
\end{align}
where
\begin{align}
&& W^{\pm}_\mu=\frac{{A}_\mu^1\pm i{A}_\mu^2}{\sqrt{2}},&&m_{\text{\tiny{$W$}}}^2=16g^2\mu^2.&&\label{4.31}
\end{align}
Fermion mass term is
\begin{align*}
{\bar\psi}<{\mit\Phi}>\psi=&g_{\text{\tiny{$Y$}}}\mu{\bar\psi}
\begin{pmatrix}
1^4 & 0^4\\
0^4 & (-1)^4
\end{pmatrix}
\psi
\end{align*}
Thus,
\begin{align}
{\cal L}_{fm}
=&m({\bar u}^ru^r+{\bar u}^gu^g+{\bar u}^bu^b+{\bar \nu}\nu)
+m({\bar d}^rd^r+{\bar d}^gd^g+{\bar d}^bd^b+{\bar e}e)
\end{align}
where we performed the transformation $e^{i\frac{\pi}{2}\gamma^5}\psi$ for 
$d^r,\,d^g,\,d^b$ and $e$ in order to make masses of these fermions have right sign. Fermion masses are all same including neutrino, so this model is really standard-like model.
\par
Let us define  photon field $A_\mu$ and the weak boson $Z_\mu$ as
\begin{align}
&&A_\mu=\frac{g'{A}^3_\mu+gB_\mu}{\sqrt{g^2+{g'}^2}},
&&
Z_\mu=\frac{g{A}^3_\mu-g'B_\mu}{\sqrt{g^2+{g'}^2}}.&&
\label{4.33}
\end{align}
Then, we can investigate the fermion electric charge which follows from
\eqref{4.17}, \eqref{4.20} and \eqref{4.33} and obtain the  charge assignment
\be
\left(\frac23, \frac23, \frac23, 0, -\frac13, -\frac13, -\frac13, -1\right)
\ee
for the fermion configuration \eqref{4.17}.\par
The standard-like model on NC spacetime proposed in this section presents 
favorable aspects of the standard model. However, it contains several 
defaults such as the UV/IR mixing and extra U(1) gauge bosons
 which exist in all gauge theories on NC spacetime. We discuss these points
in the following section.

\section{Conclusions and Discussions}
We have proposed nonabelian $\text{U}(n\otimes m)\ast$ gauge theory constructed from the commutation relations of generators  on NC spacetime.
According to this noncommutative gauge theory, 
we considered  the SU(4)$\ast$ gauge theory 
which spontaneously breaks down to $\text{U}(3\otimes 1)\ast$ symmetry
in order to obtain the gauge theory with fractional U(1) charges. 
It is shown that such NC gauge theory 
with fractional charges more than two  
can not exist alone, but it must coexist with 
NC nonabelian gauge theory.
Then, we reconstructed the standard-like model based on the gauge group
${\text{U}(3_c\otimes 2\otimes 1_{\text{\scriptsize$Y$}})\ast}$ which
 shows that fields with color as well as flavor
quantum numbers can be expressed in terms of the fundamental 
representation and also fields with fractional charges are  incorporated
in the gauge field on NC spacetime.
It also shows favorable aspects of the standard model as well as the several
defects discussed below.
\par
Let us discuss the present situations of NC gauge theories 
including models proposed here
and show 
several undesirable results with respect to the quantized version 
of NC gauge theory. 
\begin{namelist}{　\quad}
\item[　(1)]
Hayakawa \cite{Hayakawa} indicated that the charges of matter fields are restricted to $0$ and $\pm$.
\item[　(2)]
Fields can belong to the fundamental, bifundamental and adjoint representations of gauge groups \cite{GM,Tera,Chai}.
\item[　(3)]
The gauge groups are restricted to $\text{U}(N)\ast$ group \cite{Matsubara,Armoni}.
\item[　(4)]
Matusis, Susskind and Toumbas \cite{MST} found the unfamiliar IR/UV connection
in NC gauge theory which involves non-analytic behavior in NC parameter
$\theta$ making the limit $\theta\to0$ singular.
\item[　(5)]
Gauge anomalies cannot be cancelled in a chiral noncommutative theory, hence 
the anomaly free theory must be vector like \cite{Hayakawa,GM, BST}.
\end{namelist}

In this paper, we  constructed the standard-like model in order to overcome the problems (1)  and (2). There is no problem of (5) since our standard-like model is vector like.
However, the extra $\text{U}(1)$ gauge fields exist and also
our model suffers from (3) and (4) though
we do not investigate the quantum effects of the $\text{U}(n+m)\ast$
gauge theory. 
According to \cite{Khoze}, there are the following observations with respect to
 (3) and (4). In the supersymmetric version of $\text{U}(N)\ast$ gauge theory on NC spacetime \cite{MST,Khoze2,Holl}, the UV/IR mixing occurs only for the U(1) degree of freedom,  which yields the decoupling from the remaining SU$(N)$ sector at the low energy. Thus, it looks like a safe commutative SU$(N)\ast$ gauge theory at low energy. Armoni \cite{Armoni} observed the similar decoupling 
in the calculation of the one-loop gluon propagator in noncommutative QCD.
Ruiz Ruiz \cite{Ruiz} also showed that
the defects stated above may be solved 
by considering the supersymmetric version
of NC gauge theory. 
Thus, there are possibilities to be able to solve the defects stated above 
in the supersymmetric  NC gauge theory.
\par
The deviations from the standard model in particle physics have not yet been 
observed, and so any model beyond standard model must reduce to the standard model in its characteristic approximation. Then, according to the above indication, the supersymmetric gauge theory might overcome the defects stated above.
With respect to the problem (5), we have to extend the model into the left-right symmetric gauge theory.
This work will appear in future.


\end{document}